\documentclass[aps,prb,twocolumn,superscriptaddress,showpacs]{revtex4-1}

\usepackage[version=3]{mhchem}
\usepackage{graphicx}
\usepackage{siunitx}

\usepackage{hyperref}

\usepackage{url}

\usepackage{color,soul}

\usepackage{booktabs}
\usepackage[draft]{changes}
\usepackage{ctable}

\newcommand{\SiConc}{\ce{Si} concentration}

\newcommand{\magneton}{\mu_B}
\newcommand{\dgo}{\ce{Dy2Ge2O7}}
\newcommand{\dso}{\ce{Dy2Sn2O7}}
\newcommand{\dsio}{\ce{Dy2Si2O7}}
\newcommand{\dto}{\ce{Dy2Ti2O7}}
\newcommand{\hto}{\ce{Ho2Ti2O7}}

\newcommand{\hgo}{\ce{Ho2Ge2O7}}

\newcommand{\dgsoxx}{\ce{Dy2Ge_{2-x}Si_xO7}}
\newcommand{\dgsoi}{\ce{Dy2Ge_{1.98}Si_{0.02}O7}}
\newcommand{\dgsoii}{\ce{Dy2Ge_{1.92}Si_{0.08}O7}}
\newcommand{\dgsoiii}{\ce{Dy2Ge_{1.875}Si_{0.125}O7}}
\newcommand{\Fdm}{\emph{Fd}$\mathit{\bar{3}}$\emph{m}}
\newcommand{\nzo}{\ce{Nd2Zr2O7}}
\newcommand{\eg}{e.g.}

\DeclareSIQualifier{\dys}{Dy}

\DeclareSIUnit{\entropy}{\joule \per \mole \dys  \per\kelvin }

\begin{document}

\renewcommand{\figureautorefname}{Fig.}


\title{Tuning the interactions in the spin-ice materials \dgsoxx{} by
	silicon substitution}


\author{T. St\"oter}
\affiliation{Institut f\"ur Festk\"orper- und Materialphysik, TU Dresden, 01062 Dresden, Germany}
\affiliation{Dresden High Magnetic Field Laboratory (HLD-EMFL), Helmholtz-Zentrum Dresden-Rossendorf, 01328 Dresden, Germany}

\author{M. Antlauf}
\affiliation{Institut f\"ur Anorganische Chemie, TU Bergakademie Freiberg, 09596 Freiberg, Germany}

\author{L. Opherden}
\affiliation{Institut f\"ur Festk\"orper- und Materialphysik, TU Dresden, 01062 Dresden, Germany}
\affiliation{Dresden High Magnetic Field Laboratory (HLD-EMFL), Helmholtz-Zentrum Dresden-Rossendorf, 01328 Dresden, Germany}

\author{T. Gottschall}
\affiliation{Dresden High Magnetic Field Laboratory (HLD-EMFL), Helmholtz-Zentrum Dresden-Rossendorf, 01328 Dresden, Germany}

\author{J. Hornung}
\affiliation{Institut f\"ur Festk\"orper- und Materialphysik, TU Dresden, 01062 Dresden, Germany}
\affiliation{Dresden High Magnetic Field Laboratory (HLD-EMFL), Helmholtz-Zentrum Dresden-Rossendorf, 01328 Dresden, Germany}

\author{J. Gronemann}
\affiliation{Institut f\"ur Festk\"orper- und Materialphysik, TU Dresden, 01062 Dresden, Germany}
\affiliation{Dresden High Magnetic Field Laboratory (HLD-EMFL), Helmholtz-Zentrum Dresden-Rossendorf, 01328 Dresden, Germany}

\author{T. Herrmannsd\"orfer}
\affiliation{Dresden High Magnetic Field Laboratory (HLD-EMFL), Helmholtz-Zentrum Dresden-Rossendorf, 01328 Dresden, Germany}

\author{S. Granovsky}
\affiliation{Institut f\"ur Festk\"orper- und Materialphysik, TU Dresden, 01062 Dresden, Germany}
\affiliation{Faculty of Physics, M. V. Lomonossov Moscow State University, 119991 Moscow, Russia}

\author{M. Schwarz}
\affiliation{Institut f\"ur Anorganische Chemie, TU Bergakademie Freiberg, 09596 Freiberg, Germany}

\author{M. Doerr}
\affiliation{Institut f\"ur Festk\"orper- und Materialphysik, TU Dresden, 01062 Dresden, Germany}

\author{H.-H. Klauss}
\affiliation{Institut f\"ur Festk\"orper- und Materialphysik, TU Dresden, 01062 Dresden, Germany}

\author{E. Kroke}
\affiliation{Institut f\"ur Anorganische Chemie, TU Bergakademie Freiberg, 09596 Freiberg, Germany}

\author{J. Wosnitza}
\affiliation{Institut f\"ur Festk\"orper- und Materialphysik, TU Dresden, 01062 Dresden, Germany}
\affiliation{Dresden High Magnetic Field Laboratory (HLD-EMFL), Helmholtz-Zentrum Dresden-Rossendorf, 01328 Dresden, Germany}


\date{\today}

\begin{abstract}

We report that the lattice constant of \dgsoxx{} ($x=0, 0.02, 0.08, 0.125$) can
	be systematically reduced by substituting the non-magnetic germanium
	ion in the cubic pyrochlore oxide with silicon.
A multi-anvil high-pressure synthesis was performed up to \SI{16}{\giga \pascal}
	and \SI{1100}{\celsius} to obtain polycrystalline samples in a
	solid-state reaction.
Measurements of magnetization, ac susceptibility, and heat capacity reveal the
	typical signatures of a spin-ice phase.
From the temperature shift of the peaks, observed in the temperature-dependent
	heat capacity, we deduce an increase of the strength of the exchange
	interaction.
In conclusion, the reduced lattice constant leads to a changed ratio of the
	competing exchange and dipolar interaction.
This puts the new spin-ice compounds closer towards the phase boundary of
	short-range spin-ice arrangement and antiferromagnetic long-range order
	consistent with an observed reduction of the energy scale of monopole
	excitations.
\end{abstract}


\maketitle

\section{Introduction}

Magnetically frustrated materials have attracted
	considerable interest because of their rich magnetic phases hosting
	exotic states of matter and emergent excitations%
	\citep{greedan_geometrically_2001, greedan_frustrated_2006}.
Frustration in a magnetic material can lead to a ground-state
	manifold without long-range order\citep{greedan_geometrically_2001}.
This can occur in various possible ways, such as via antiferromagnetic (AFM)
	nearest-neighbor coupling of Ising spins on a triangular lattice or a
	periodical variation of magnetic interactions between nearest
	neighbors\citep{ramirez_strongly_1994, wosnitza_frustrated_2016}.

\begin{figure}
	\centering
	\includegraphics[width=0.8\linewidth]{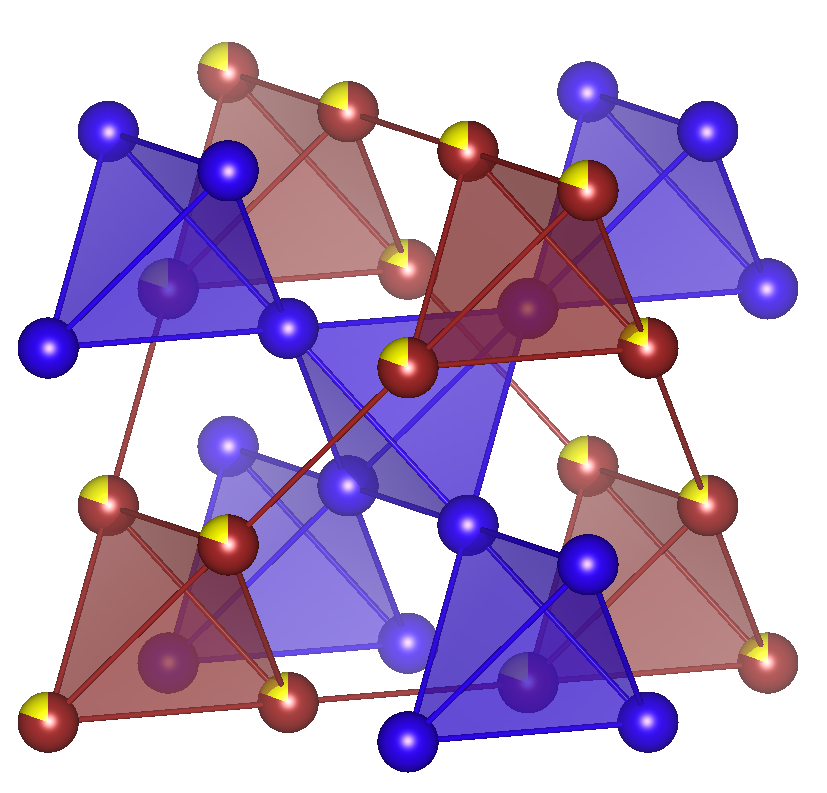}
	\caption{Structure of \dgsoxx{}: the $A$ site is occupied by
		\ce{Dy^3+} (shown as blue spheres) and the $B$ site is
		partially occupied by \ce{Si^4+} and \ce{Ge^4+} shown as red
		spheres with a yellow segment; the oxygen atoms are omitted.
		\label{fig:structure}
		}
\end{figure}
\begin{figure}
	\centering
	\includegraphics[width=\linewidth]{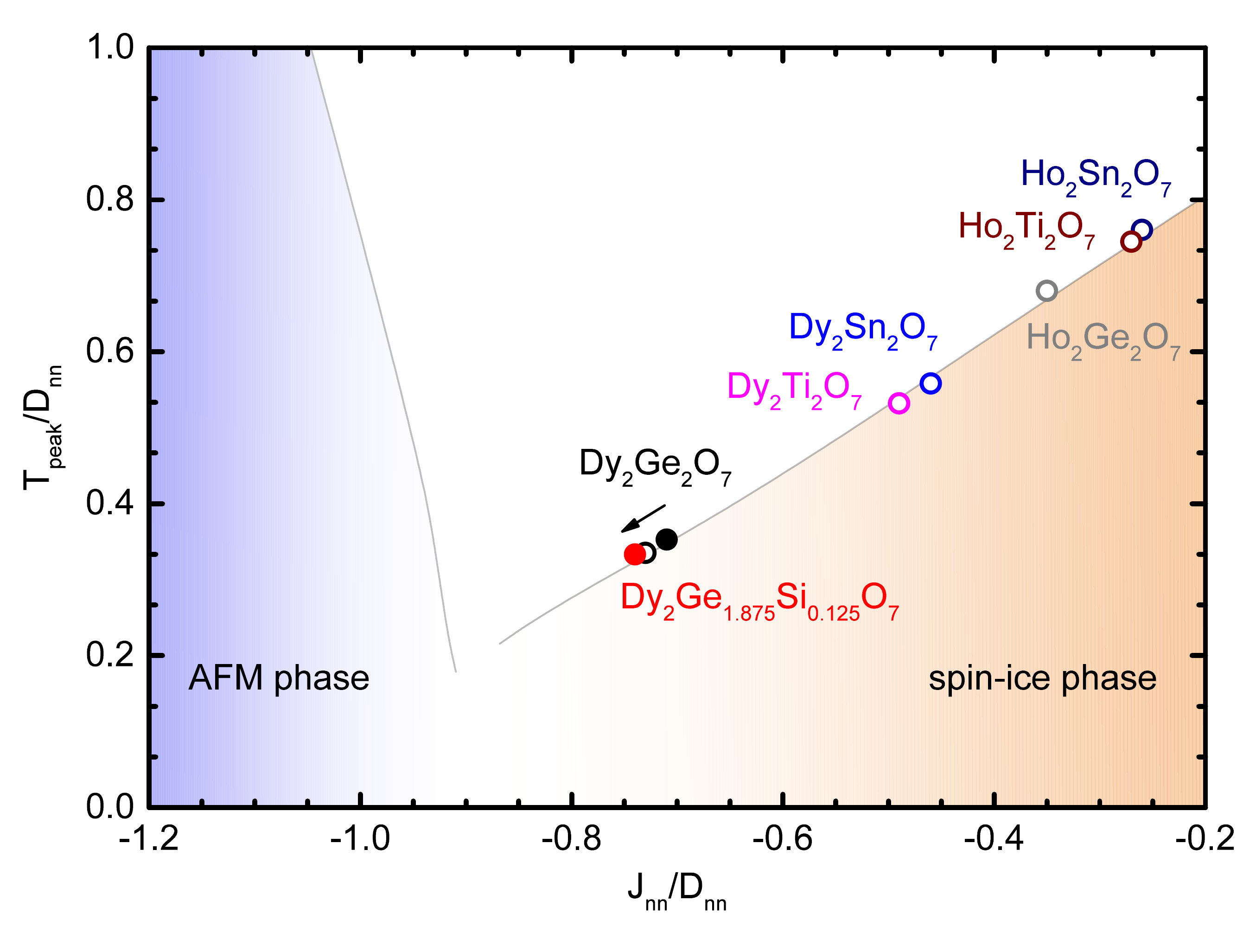}
	\caption{Phase diagram of the dipolar spin-ice
		model{\citep{den_hertog_dipolar_2000}} including the
		positions of several spin-ice compounds and the phase boundaries
		of the antiferromagnetic (AFM) and the spin-ice phase.
		Data from literature{\citep{zhou_chemical_2012}} are shown
		as open circles, filled symbols
		stand for the samples from this work.
		$J_\mathrm{nn}$ and $D_\mathrm{nn}$ denote the exchange and dipolar
		interaction between nearest neighbors, respectively.
		$T_\mathrm{peak}$ denotes the temperature of the peak in the
		heat capacity associated with the spin-ice phase (adapted
		from~\citet{zhou_chemical_2012}).
		\label{fig:dsi_phase_diagram}
		}
\end{figure}

In the case of pyrochlore oxides \ce{$A$2$B$2O7} with a trivalent ion
	\ce{$A$^3+} and a tetravalent ion \ce{$B$^4+} (\autoref{fig:structure}),
	frustration can arise due to the strong crystal electric field acting on
	the ions on the pyrochlore lattice of corner-sharing tetrahedra.
For pyrochlores with a rare-earth ion on the $A$ site many different exotic
	states have been observed, such as classical or quantum spin-ice states
	with magnetic-monopole excitation%
	\cite{balents_spin_2010, gingras_quantum_2014, zvyagin_new_2013,
	gardner_magnetic_2010, greedan_frustrated_2006,rau_frustrated_2019}.
Specifically, for the pyrochlores with \ce{Dy} and \ce{Ho} on the $A$ site, the
	rare-earth ion is subject to a strong crystal-electric-field splitting
	of the ground state and the first excited state of the single ion by an
	energy of the order of \SI{200}{\kelvin}%
	\citep{rosenkranz_crystal-field_2000}.
The single-ion ground state is an effective spin-half state%
	\citep{rau_magnitude_2015}
	pointing either in or out of the tetrahedra the ion belongs to.
The two-ion interaction is well described by the dipolar spin-ice model
	which includes dipolar $D_\mathrm{nn}$ and exchange interaction
	$J_\mathrm{nn}$ to form an effective interaction
	$J_\mathrm{eff} = D_\mathrm{nn} + J_\mathrm{nn}$ between nearest
	neighbors{\citep{den_hertog_dipolar_2000}}.
Following this simple spin-ice model{\citep{den_hertog_dipolar_2000}} for
	{\hto{}} and {\dto{}}, the
	competing interactions result in an effective ferromagnetic
	nearest-neighbor interaction $J_\mathrm{eff} > 0$, and a
	positioning of these compounds on the right-hand side
	of the phase diagram of this model {\autoref{fig:dsi_phase_diagram}} --
	in contrast, neodymium-based pyrochlores
	($e.g.$ \nzo{}) adopt an AFM long-range ordered phase situated on the
	left-hand side of {\autoref{fig:dsi_phase_diagram}}%
	{\citep{Bertin_2015, Opherden_2017,opherden_inverted_2018,
	xu_magnetic_2015}}
	because of a different relation of exchange to dipolar interaction.
An effective interaction $J_\mathrm{eff} > 0$ favors the highly degenerate
	spin-ice configuration:
	two spins point into and two spins point out of each tetrahedron
	(``2-in-2-out''){\citep{bramwell_spin_2001}}.
The excitations of this arrangement can be interpreted as the creation of
	magnetic monopole-antimonopole pairs\cite{castelnovo_magnetic_2008}.
In several investigations of {\dto{}} and {\hto{}} via ac-susceptibility
	measurements{\citep{matsuhira_novel_2001, snyder_quantum-classical_2003,
	snyder_low-temperature_2004, matsuhira_spin_2011,
	yaraskavitch_spin_2012, matsuhira_low_2000,
	snyder_how_2001, ehlers_dynamical_2003, ehlers_evidence_2004,
	quilliam_dynamics_2011}} thermally-activated spin dynamics have been
	found that are attributed to the thermal excitation of
	monopoles{\citep{jaubert_magnetic_2011}}.
The magnetization and specific-heat measurements of the spin-ices
	show characteristic behavior like a liquid-gas transition at sub-kelvin
	temperatures in field-dependent measurements%
	{\citep{sakakibara_observation_2003, krey_first_2012,
	petrenko_magnetization_2003}}
	or a Schottky-like anomaly in temperature-dependent measurements at
	zero field{\citep{hiroi_specific_2003,
	higashinaka_low_2004,bramwell_spin_correlation_2001}}, respectively.
A strong experimental evidence for the dipolar spin-ice character is the
	residual Pauling entropy of $S = 1/2 \ln{(3/2)}R$ calculated from
	measurements of the specific heat%
	{\citep{ramirez_zero-point_1999, den_hertog_dipolar_2000,
	giblin_pauling_2018}}.

Although recent publications{\citep{pomaranski_absence_2013,
	giblin_pauling_2018}} discuss
	deviations from this Pauling's ice-rule entropy due to extremely slow
	relaxation dynamics below {\SI{1}{\kelvin}}, a non-vanishing
	entropy on shorter time scales remains a prerequisite of the spin-ice
	state{\citep{giblin_pauling_2018}}.
Theoretical investigations of spin-ice models including interactions beyond
	nearest neighbors find a long-range ordered ground state satisfying the
	ice-rules{\citep{melko_long-range_2001,melko_monte_2004,
		mcclarty_chain-based_2015,yavorskii_dy_2008,
		henelius_refrustration_2016,rau_spin_2016}}.
Comparing our samples with these refined models lies beyond our experimental
	possibilities and we concentrate on the dependence of the
	nearest-neighbor exchange constant $J_\mathrm{eff}$ on the {\ce{Si}}
	substitution $x$ in {\dgsoxx{}}.

Especially in the vicinity of the transition between long-range (AFM) and
	short-range (spin-ice) correlations,
	new magnetic states
	may occur%
	{\citep{zhou_high_2011, wiebe_frustration_2015}}.
Therefore, tuning the internal interactions seems a promising route to
	investigate these new states within this domain of the phase space.
Instead of the experimentally difficult application of hydrostatic
	pressure to existing sample
	that manipulates the nearest-neighbor distance%
	{\citep{mirebeau_spin_2004}}, the interaction between neighboring
	rare-earth ions can be varied by
	exchanging the $B$ site ion with one of different ionic
	radius\citep{zhou_chemical_2012, wiebe_frustration_2015}.
For example, substituting \ce{Ge} for \ce{Ti} in \dto{} reduces the distance
	between the \ce{Dy} ions and thus changing the relative strengths of the
	competing dipolar and exchange interactions, $D_\mathrm{nn}$ and
	$J_\mathrm{nn}$, respectively\citep{zhou_chemical_2012}.
The total effective interaction $J_\mathrm{eff}$
	is reduced shifting the compound closer to the not yet
	experimentally investigated
	phase region just between the spin-ice and AFM phase in the dipolar
	spin-ice model, see \autoref{fig:dsi_phase_diagram}.

The next challenge is to synthesize a material with an even smaller $B$ site
	ion than \ce{Ge}, such as \ce{Si}, allowing for further tuning the
	balance of exchange ($J_\mathrm{nn}$) and dipolar interactions
	($D_\mathrm{nn}$) in a spin-ice compound.
This new ratio $J_\mathrm{nn}/D_\mathrm{nn}$ allows for investigating
	pyrochlores closer to the crossover region between frustration-driven
	dynamics and magnetic order than ever before.

But many combinations of $A$ and $B$ site ions have an unstable pyrochlore phase
	when synthesized at ambient pressure\citep{mouta_tolerance_2013}.
\citet{mouta_tolerance_2013} have empirically investigated the stability of the
	pyrochlore structure with respect to the ratio of the atomic radii
	$r_{a}/r_{b}$, where $r_a$ and $r_b$ denote the atomic radii
	of the $A$ site and $B$ site ions, respectively.
At ambient pressure, stable pyrochlores can be realized for ratios
	$1.36 < r_{a}/r_{b} < 1.71$%
	\citep{wiebe_frustration_2015, gardner_magnetic_2010}.
For comparison, the ratio of the ionic radii for \ce{Dy^3+} on the
	$A$ site and \ce{Ti^4+}, \ce{Ge^4+}, and \ce{Si^4+} on the $B$ site are
	$1.70$, $1.94$, and $2.57$, respectively.
Under high hydrostatic pressure during synthesis, the stability range can be
	extended towards smaller $r_b$, \eg{} from \dto{} to \dgo{}.
The resulting pyrochlores persist metastable also under ambient pressure
	conditions\citep{wiebe_frustration_2015}.

While for the growth of \dsio{} with the hypothetical ratio $r_a/r_b = 2.57$,
	pressures beyond the state of the art would be needed or this phase
	would be inherently unstable, we were able to grow stable \dgsoxx{}
	polycrystals with \ce{Si} substitution of up to $x = 0.125$ under
	high-pressure and high-temperature conditions.
In this paper, we show that such a substituted compound exhibits the
	characteristic properties
	of a classical spin-ice material while having reduced effective
	interactions that shift its magnetic state further towards the
	phase
	boundary between spin-ice and AFM order.

\section{Experimental}

Polycrystalline \ce{Dy} germanate \ce{Dy2Ge2O7} was synthesized similar as
	described previously\cite{zhou_high_2011,shannon_effective_1969}:
\ce{GeO2} (Heraeus, \SI{99,999}{\percent}) and \ce{Dy2O3}
	(Chempur, \SI{99,999}{\percent}) were used as starting materials.
Stoichiometric amounts were mixed with ethanol and thoroughly ground in an
	agate mortar until dryness.
The synthesis was performed in a ``Walker-type'' high-pressure apparatus%
	\cite{walker_simplifications_1990} with a \SI{1000}{\tonne} hydraulic
	press (Voggenreiter GmbH).
The powder mixture was filled in an $h$-BN capsule and compressed in an
	\SI{18}{\milli \meter} zirconia octahedron pressure medium with a load
	of \SI{970}{\tonne} corresponding to a sample pressure of
	\SI{8}{\giga \pascal}.
The pyrochlore material forms during a solid-state reaction at
	\SI{1100}{\degreeCelsius} maintained for \SI{90}{\minute}.
The temperature was measured with a Type C (\ce{W}5\%\ce{Re}/\ce{W}26\%\ce{Re})
	thermocouple.

For the polycrystalline silicone-substituted \dgsoxx{} samples, \ce{SiO2}
	(Degussa Aerosil380, $>\SI{99,8}{\percent}$) was used additionally.
The uni-axially prepressed samples were surrounded by \ce{CsCl}
	capsules and compressed in \SI{10}{\milli \meter} \ce{MgO}:\ce{Cr}
	octahedral pressure cells (Ceramic Substrate \& Components Ltd.) with a
	load of \SI{642}{\tonne} which equals a sample pressure of approximately
	\SI{16}{\giga \pascal}.
\ce{Si}-containing pyrochlores were synthesized at \SI{1100}{\degreeCelsius}
	for 90 to \SI{180}{\minute}.
We recovered polycrystalline, sintered
	pieces of dark to light gray material after the high-pressure
	synthesis of \dgsoxx{} with $x = 0, 0.02, 0.08$, and $0.125$.
The phase purity was confirmed via SEM/EDX analysis (Carl Zeiss LEO
	1530) for all samples and via room-temperature x-ray powder
	diffraction at the ALBA synchrotron light source (Beamline 4,
	$\lambda = \SI{0.413364}{\angstrom}$) in
	Barcelona, Spain, for \dgo{} and by using a Seifert diffractometer
	(FPM URD6) in symmetric Bragg-Brentano-Geometry with \ce{Cu} anode for
	\dgsoxx{} with $x=0.02,0.08$, and $0.125$.
The software package Fullprof (Suite 3.00, 2015) was used for the Rietveld
	analysis to obtain the lattice constants.

Magnetization measurements were performed using a commercial SQUID magnetometer
	(MPMS) and a vibrating-sample magnetometer in magnetic fields up to
	\SI{7}{\tesla}.
Magnetic ac-susceptibility measurements were performed in a compensated
	coil-pair susceptometer at frequencies ranging from \SI{4}{} to
	\SI{1293}{\hertz} down to a temperature of \SI{0.3}{\kelvin} using a
	commercial $^3$He system.
The temperature was measured with a \ce{RuO2} resistance thermometer.
Demagnetization effects{\citep{quilliam_dynamics_2011}} were estimated and
	found to be of no importance for our arguments.
The heat capacity of \dgo{} and \dgsoiii{} was measured
	using the quasi-adiabatic heat-pulse method{\citep{barron_heat_1999}}
	in which the sample is heated by a short pulse of defined energy while
	the resulting temperature change is measured.
For this the sample is placed on a sapphire platform in a $^3$He
	sorb-pumped cryostat with a carefully calibrated {\ce{RuO2}} thermometer.

\section{Results and Discussion}

\begin{figure}
	\centering
	\includegraphics[width=\linewidth]{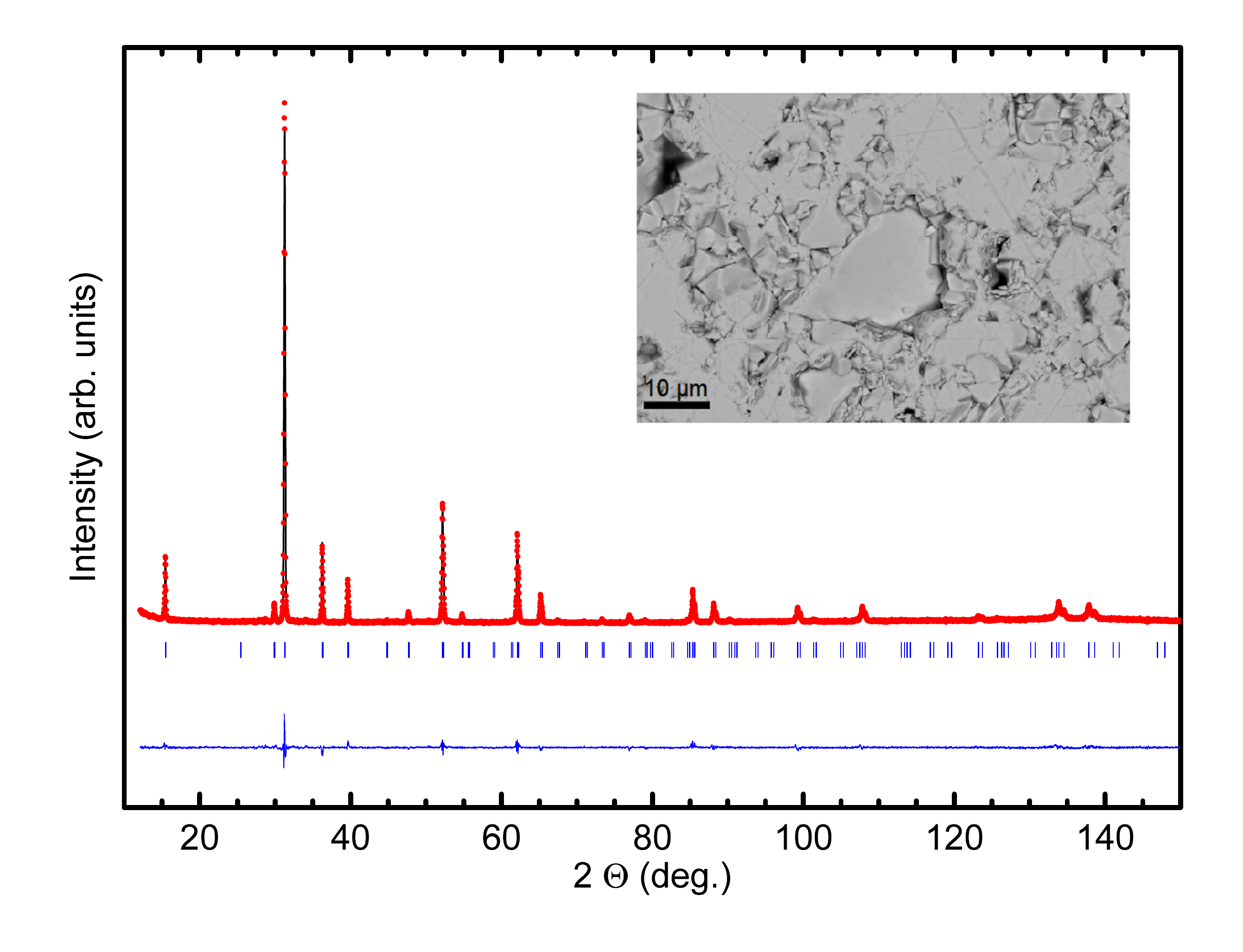}
	\caption{Powder-diffraction pattern and Rietveld-difference
		plot for \dgsoxx{} with $x = 0.125$, fitted for this
		stoichiometry.
		Inset: SEM micrograph of the polished sample surface,
		revealing crystallite sizes up to approximately
		\SI{15}{\micro \meter}.
		\label{fig:XRD}
		}
\end{figure}
The diffraction data of unsubstituted Dy germanate confirmed the cubic
	pyrochlore structure (\Fdm, 227) with a lattice constant of
	$a= \SI{9,930+-0.001}{\angstrom}$, which is in excellent agreement with
	the literature data ($a= \SI{9,929}{\angstrom}$\cite{zhou_high_2011}).
Rietveld refinements show that the pyrochlore structure can well describe
	the diffraction patterns of \dgsoi{}, \dgsoii{}, and \dgsoiii{}
	(the latter is shown in \autoref{fig:XRD}) with lattice constants
	$a = \SI{9.924+-0.001}{\angstrom}$, $\SI{9.912+-0.001}{\angstrom}$, and
	$\SI{9,906+-0.001}{\angstrom}$, respectively.
We achieved the best fit by assuming a statistical substitution of \ce{Ge} with
	\ce{Si} on the $B$ site of the pyrochlore structure with exactly
	the \SiConc{} of the starting-material mixtures.
In particular, \ce{Si} is not interstitially incorporated into the
	crystal.
There were no signs for chemical inhomogeneities or additional phases in XRD or
	SEM investigations (\autoref{fig:XRD}).

\begin{figure}
	\centering
	\includegraphics[width=\linewidth]{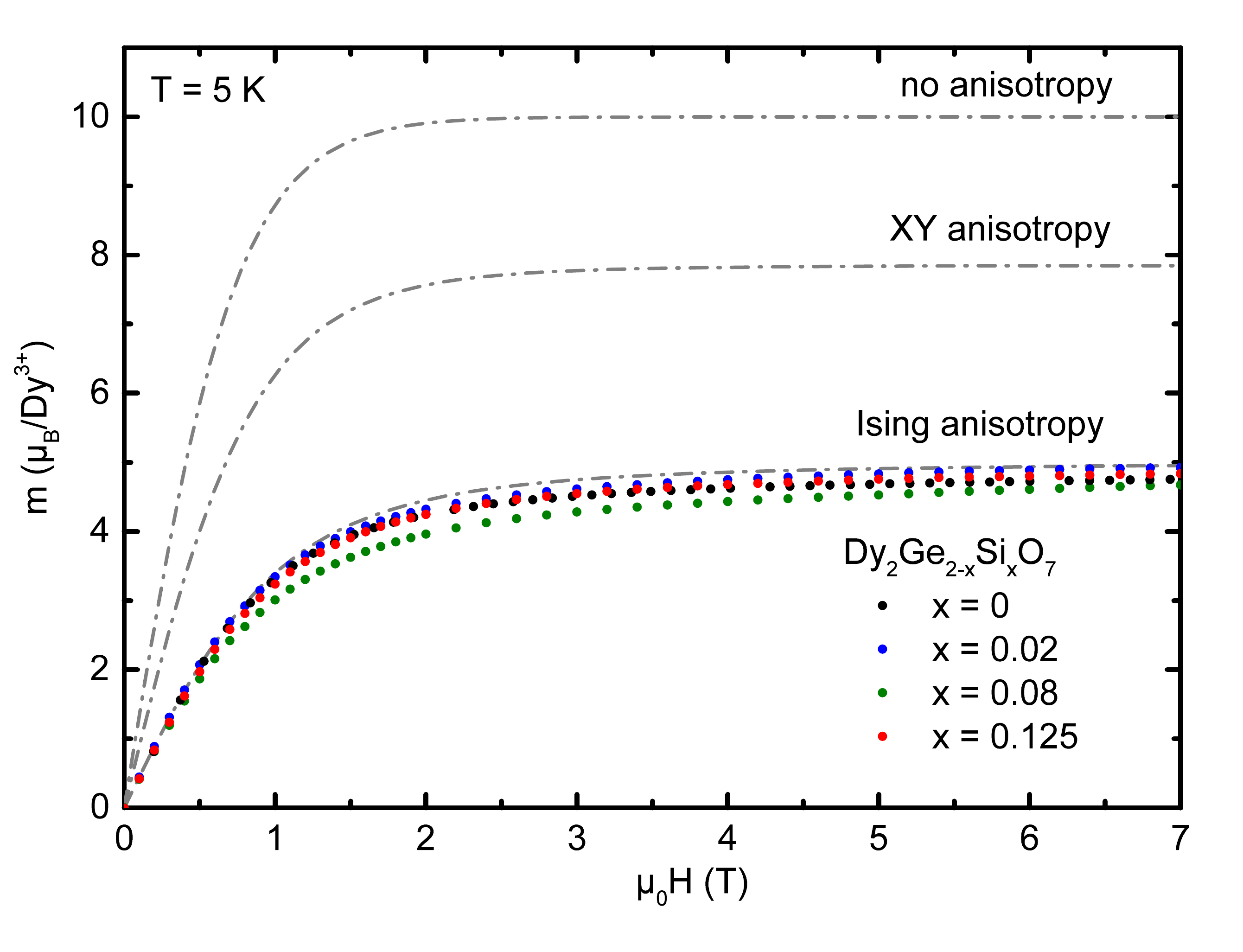}
	\caption{Field dependence of the magnetization per dysprosium ion of
		\dgsoxx{} with $x = 0,0.02,0.08$, and $0.125$ at a temperature
		of \SI{5}{\kelvin}.
		The symbols are measured values and the dash-dotted lines
		correspond to the powder-averaged Boltzmann distribution of
		spin-half spins with Ising-anisotropy, XY-anisotropy, and
		with no anisotropy, respectively.
		The error bars are within the size of the symbols.
		\label{fig:magnetization_all}
		}
\end{figure}
In \autoref{fig:magnetization_all}, the magnetization per \ce{Dy} ion of the
	sample series is shown as a function of magnetic field measured at low
	temperatures and consistent with the
	literature{\citep{bramwell_bulk_2000}}.
The field dependence of the magnetization is well described by a powder-averaged
	Boltzmann distribution of non-interacting paramagnetic Ising spins
	(dash-dotted curve) resulting in a magnetic moment of
	$\SI{4.8+-.15}{\magneton}$ per \ce{Dy} at saturation while the
	Heisenberg model (no anisotropy) or the case of XY-anisotropy would
	result in a very different magnetic behavior.
It should be noted that the magnetization values at the maximum field do
	not systematically depend on the {\ce{Si}} concentration.
The deviations 
	can be attributed to the use
	of polycrystalline samples which, due to the growing process, may
	contain a degree of texture which is not visible in the diffractograms.
However, the low scattering of the measuring points shows the accuracy of the
	magnetization measurements.
For the calculation of the $D_\mathrm{nn}$ the value of {\SI{10}{\mu_B}} for the
	free moment was used.
The Curie-Weiss temperature {$\theta_\mathrm{CW}$} of our {\dgo{}} sample
	is close to zero and consistent with the literature value%
	{\citep{zhou_chemical_2012}}; the difference of {$\theta_\mathrm{CW}$}
	of the {\ce{Si}} containing samples compared to {\dgo{}} is within the
	accuracy of the measurement of approximately {\SI{\pm 0.5}{\kelvin}}.

\begin{figure*}
	\centering
	\includegraphics[width=\textwidth]{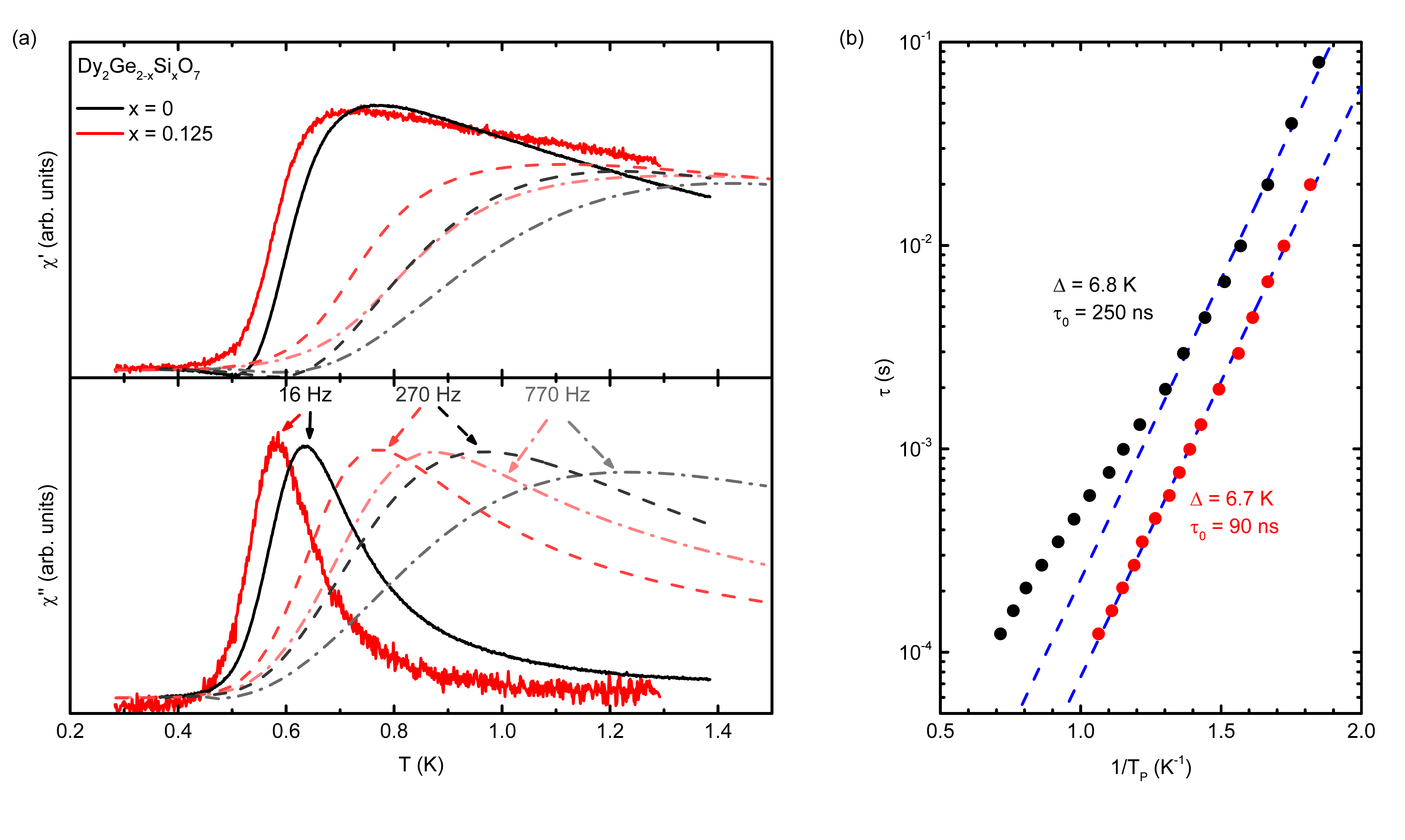}
	\caption{(a) Real and imaginary part of the ac susceptibility $\chi'$
		and $\chi''$ of \dgo{} (black) and \dgsoiii{} (red) at
		frequencies of \SI{16}{} (solid), \SI{270}{} (dashed), and
		\SI{770}{\hertz} (dash-dotted).
		(b) Spin-relaxation time $\tau = (2\pi f)^{-1}$ as a function
		of the inverse peak temperature $T_P$.
		Fits using the Arrhenius law with their respective energy
		barriers $\Delta$ are shown as dashed lines.
		The data are not corrected for demagnetization.
		\label{fig:ac_susceptibility_phase_diagram}
		}
\end{figure*}

To detect distinct differences, we will focus our investigation for the
	remainder of this article on pure Dy germanate and the sample with the
	highest \ce{Si} substitution, \dgsoiii{}.
Figure~\ref{fig:ac_susceptibility_phase_diagram}(a) shows the real ($\chi'$)
	and imaginary part ($\chi''$) of the ac susceptibility of the
	substituted and unsubstituted sample for three exemplary frequencies.
The distinct frequency dependence is visible especially in $\chi''$.
Both samples show sharp maxima which shift towards higher temperatures and
	smear out continuously at higher frequencies.
At \SI{16}{\hertz}, \dgsoiii{} has a peak temperature of \SI{580}{\milli\kelvin}
	being about \SI{50}{\milli\kelvin} lower than for \dgo{}.
Furthermore, it is observed that the increase of the peak temperature at higher
	frequencies is much less pronounced in the substituted compound.
Demagnetization effects should not influence our following qualitative
	discussion, even though they can cause deviations of the measured
	susceptibility, for the absolute values as well as for the
	peak positions{\citep{quilliam_dynamics_2011,yaraskavitch_spin_2012}}.
They should cause similar changes for the equally shaped \dgo{} and
	\dgsoiii{} samples; however, an analytical determination of the
	demagnetization factor for our disc-shaped samples is not possible.

Based on the ac-susceptibility results, the spin dynamics can be investigated
	further.
At the peak temperature $T_\mathrm{P}$ of the imaginary part $\chi''$ at a
	given attempt frequency $f$, the spin-relaxation time $\tau$ is related
	to $f$ via $\tau = 1/2\pi f$.
In Fig.~\ref{fig:ac_susceptibility_phase_diagram}(b), the spin-relaxation time
	is plotted vs.\ the inverse of the temperature $T_\mathrm{P}$.
{\dgo{}} has two linear regions; a high-temperature region above about 
	{\SI{0.8}{\kelvin}} that is less steep than the low-temperature region.
In {\dgsoiii{}} only one slope is found below {\SI{1}{\kelvin}}.
We focus our discussion on the low temperature region.
At low measurement frequencies, the relaxation times seem to follow an
	Arrhenius law (dashed lines in
	Fig.~\ref{fig:ac_susceptibility_phase_diagram}(b)):
\begin{equation}\label{key}
	\tau (T_\mathrm{P}) =
		\tau_0 \exp (\Delta/k_B T_\mathrm{P}) \, .
\end{equation}

The energy barriers $\Delta$ used to fit the data change only little, from
	\SI{6.8+-.2}{\kelvin} to \SI{6.70+-.05}{\kelvin}, due to the partial
	substitution of \ce{Ge} by \ce{Si}.
Instead, the shift of the peak positions of $\chi''$ from {\dgo{}} to
	{\dgsoiii{}} mainly result in a change of the pre-exponential factor
	$\tau_0$ from {\SI{250+-50}{\nano \second}} to
	{\SI{90+-10}{\nano\second}} due to the {\ce{Si}} substitution.
This pre-exponential factor is attributed to the spin-tunneling rate between
	the two Ising states and should be determined by the systems CEF level
	scheme and the transverse fields acting on the flipping spin%
	{\citep{jaubert_magnetic_2011,tomasello_correlated_2018}}.
The reduction of $\tau_0$ with substitution of {\ce{Ge}} by {\ce{Si}} is not
	surprising, as it decreases also with decreasing lattice constant from
	{\dso{}} to {\dto{}}{\citep{matsuhira_slow_2011}}.
Due to the similar shape of the samples, demagnetization
	effects are unlikely to alter the qualitative observation of a
	reduction of
	the pre-exponential factor from {\dgo{}} to \dgsoiii{}.

Apart from the influence of the mere lattice contraction, however, the random
	distribution of {\ce{SI}} and {\ce{Ge}}
	on the $B$ site may contribute to this reduction.
Possibly, the lowering of the local site symmetry influences the Ising
	character of the single ion inducing transverse exchange coupling%
	{\citep{rau_magnitude_2015}} as perturbation of the dominant Ising
	interaction.
Such transverse exchange coupling might reduce $\tau_0$
	as proposed in other rare-earth compounds
	with spin-ice character{\citep{gao_dipolar_2018}}.

\begin{figure}
	\centering
	\includegraphics[width=\linewidth]{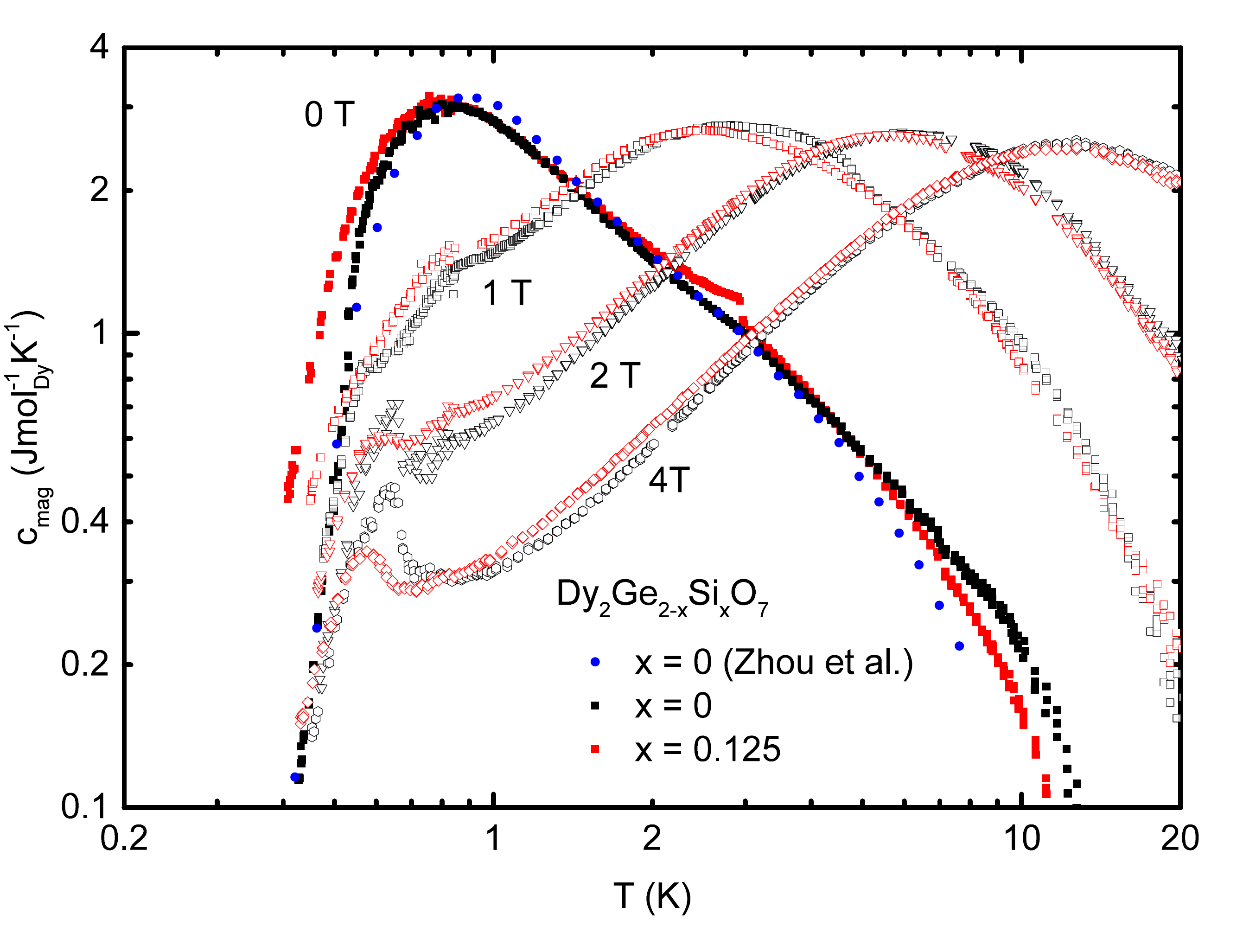}
	\caption{Temperature dependence of the magnetic specific heat
		$c_{\text{mag}}$ per mole Dy of the
		\dgo{} and \dgsoiii{} samples in black and red symbols,
		respectively.
		The zero-field data is marked by full symbols and the data at
		{\SI{1}{}}, {\SI{2}{}} and  {\SI{4}{\tesla}}
		by empty symbols.
		Specific heat data for {\dgo{}} from \citet{zhou_chemical_2012}
		(blue) have been included for comparison.
		\label{fig:DGSOHeatCapacity}
		}
\end{figure}
The magnetic specific heat $c_\mathrm{mag}$ of \dgo{} and \dgsoiii{} in
	external fields
	up to \SI{4}{\tesla} is shown in \autoref{fig:DGSOHeatCapacity}.
The unsubstituted sample has a peak in the heat capacity at a temperature of
	{\SI{0.84+-0.01}{\kelvin}},
	determined by a phenomenological fit,
	whereas the substituted compound has a
	slightly reduced peak position of {\SI{0.80+-0.01}{\kelvin}} and
	an increased peak height.
Since this work deals with polycrystalline samples, as does
	Ref.~{\citep{zhou_chemical_2012}}, the peak height can be influenced by
	factors both in production and measurement.
However, a rise of the peak height in the substituted sample fits into the
	picture of a smaller {$J_\mathrm{nn}/D_\mathrm{nn}$}
	{\citep{den_hertog_dipolar_2000}}.

This Schottky-like peak is associated to spin-freezing and establishment of
	the spin-ice state{\citep{ramirez_zero-point_1999}}.
Even though the peak temperature is shifted only by a small amount,
	it is a strong hint that the substitution of \ce{Si}
	leads to a shift towards the boundary between spin-ice and AFM phase in
	the phase diagram (\autoref{fig:dsi_phase_diagram}).
At temperatures above the peak position, both compounds show almost the same
	specific heat; above {\SI{10}{\kelvin}} the phononic contribution becomes
	dominant.
Measurements up to {\SI{30}{\kelvin}} are used to determine this contribution in
	order to extrapolate it to the low-temperature specific heat.
The magnetic specific heat is the total specific heat minus the phononic
	contribution.
The feature in the specific heat of \dgsoiii{} at about
	\SI{3}{\kelvin} could indicate the possible presence of a minority
	phase that was not visible in the XRD studies.
Furthermore, a broadening of the specific-heat peak of the substituted material
	towards lower temperatures compared to \dgo{} can be seen.
However, in an ideal spin-ice with smaller {$J_\mathrm{nn}/D_\mathrm{nn}$} than
	\dgo{} we would rather expect a narrowing of the specific-heat peak%
	{\citep{den_hertog_dipolar_2000}}.
The origin for this broadening could be related to the
	random occupation of the {\ce{Si}} and {\ce{Ge}} atoms on the $B$ site
	altering the bond environments.
A distribution of bond environments might result in a distribution of exchange
	constants broadening the Schottky-like specific-heat peak.

In magnetic fields, the peak in the specific heat is broadened and
	shifted to higher temperatures compared to zero field
	which is an expected
	behavior for this anomaly {\citep{Kim_2014,ramirez_zero-point_1999,
	Higashinaka_2002}}.
The additional features that arise in field are also observed in
	Ref.~{\citep{ramirez_zero-point_1999}} and may be 
	attributed to the polycrystalline nature
	of the samples{\citep{Higashinaka_2002}}.
Additional explanations are given in Ref.~{\citep{Ruff_2005}} by simulation
	methods.
A definite conclusion could only be drawn with data
	obtained from single crystals.

\begin{figure}
	\centering
	\includegraphics[width=\linewidth]{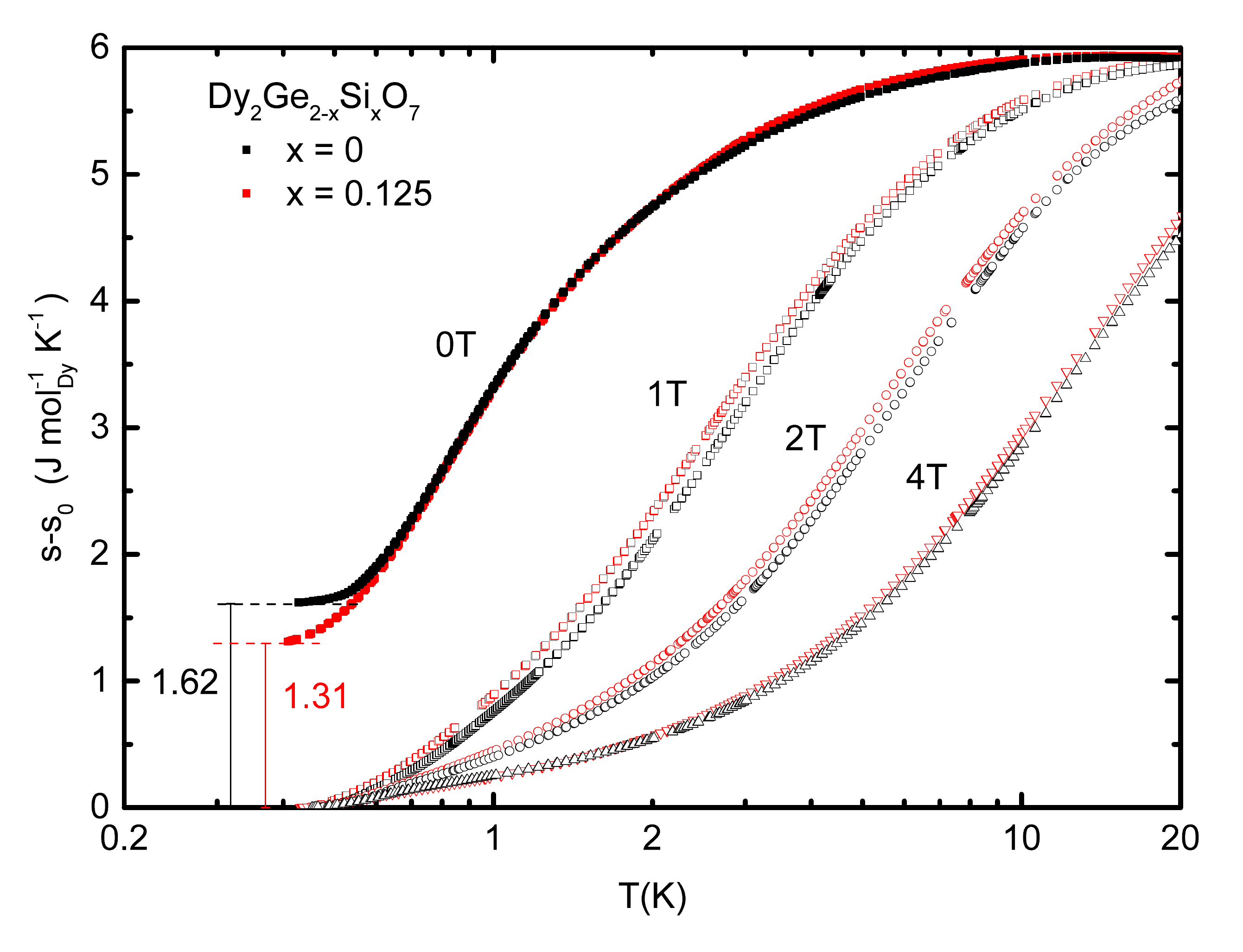}
	\caption{Temperature dependence of the molar entropy of \dgo{} (black)
	and \dgsoiii{} (red) at zero field (filled) and
	{\SI{1}{}}, {\SI{2}{}} and  {\SI{4}{\tesla}} (empty).
	\label{fig:DGSOEntropy}
	}
\end{figure}

The magnetic entropy was calculated from $c_\mathrm{mag}/T$ by integrating
	downwards from the temperature at which the curves with and without
	field overlap and fixing the plateaus at this value.
The entropy data of {\dgo{}} and {\dgsoiii{}} at zero field,
	{\SI{1}{}}, {\SI{2}{}} and {\SI{4}{\tesla}} are compared
	in Fig.~{\ref{fig:DGSOEntropy}}.
The magnetic entropy of {\dgo{}} and {\dgsoiii{}} shows a similar functional
	behavior.
In external magnetic fields of {\SI{1}{\tesla}} and higher, the entropy of
	the ground state is recovered ({\autoref{fig:DGSOEntropy}}).
This behavior in an external field is typical for a spin-ice
	{\citep{ramirez_zero-point_1999,Higashinaka_2002,bramwell_spin_2001}}
	since the external magnetic field lifts the degeneracy of the ground
	state.
The value of the ground-state entropy with {\SI{1.62}{\entropy}} (unsubstituted)
	and {\SI{1.31}{\entropy}} falls somewhat short of the Pauling entropy
	of {\SI{1.69}{\entropy}} of the ideal spin ice on a perfect crystal.
The same arguments as for the peak height and width apply for this reduction,
	as well.
	
\begin{table}
	\caption{Lattice parameters and selected magnetic parameters to insert
		\dgo{} and \dgsoiii{} in the phase diagram
		(\autoref{fig:dsi_phase_diagram}).
		\label{tab:phasediagram}}
	\begin{tabular}{ccccccc}\hline
		x & a & $D_\mathrm{nn}$
		& $c_\mathrm{mag}(T_\mathrm{peak})$ & $T_\mathrm{peak}$
		& $J_\mathrm{nn}/D_\mathrm{nn}$ & $J_\mathrm{eff}$\\
		& (\SI{}{\angstrom}) & (\SI{}{\kelvin}) 
		& (\SI{}{\entropy}) & (\SI{}{\kelvin})
		& & (\SI{}{\kelvin})\\\hline
		0 & 9.930 & 2.40 & 3.02 & 0.84 & -0.71 & 0.69\\
		0.125 & 9.906 & 2.42 & 3.07 & 0.80 & -0.74 & 0.63\\\hline
	\end{tabular}
\end{table}
To place the new material in the phase diagram, we determined
	{$c_\mathrm{mag}$} and {$T_\mathrm{peak}$} from
	{\autoref{fig:DGSOHeatCapacity}} and calculated the other values in
	{\autoref{tab:phasediagram}}
	as explained in the following.
We determined the dipolar
	interaction constant by the common estimation:
\begin{equation}
	D_\mathrm{nn} =
	\frac{5}{3} \frac{\mu_0}{4\pi} \frac{g^2 \mu^2}{r^3_\mathrm{nn}} \, ,
	\label{eq:diopl-WW}
\end{equation}
	with the moment assumed from the theoretical value of
	$g \mu = \SI{10}{\mu_\mathrm{B}}$ and
	$r_\mathrm{nn}=\sqrt{2}a/4$
	($a$ lattice constant) being the distance between two {\ce{Dy^3+}}
	ions.
Here we used the lattice constants determined at {\SI{300}{\kelvin}} from
	the x-ray diffractograms to be consistent with other publications%
	{\citep{xu_magnetic_2015,zhou_chemical_2012}}.
Measurements of the thermal expansion of singlecrystalline samples of the
	isostructural compounds \dto{} and \hto{} down to low temperatures of
	about {\SI{1}{\kelvin}} confirm that
	the assumed relations are largely retained.
The reduction of the lattice constant of the Si-substituted sample, thus, leads
	to a slight increase of $D_\mathrm{nn}$ from \SI{2.40}{} to
	\SI{2.42}{\kelvin}.
However, the increase of the strength of the exchange interaction,
	$J_\mathrm{nn}$, is more pronounced.
Following the approach firstly described by%
	\citet{den_hertog_dipolar_2000}
	for {\dto{}} and {\hto{}} and used in several publications%
	\citep{bramwell_spin_2001,zhou_chemical_2012}
	for {\ce{Dy}} and {\ce{Ho}}
	pyrochlores including {\dgo{}} and {\hgo{}}, $J_\mathrm{nn}$ can
	be graphically determined on the
	$T_\mathrm{peak}/D_\mathrm{nn}$-$J_\mathrm{nn}/D_\mathrm{nn}$ line in
	the phase diagram (\autoref{fig:dsi_phase_diagram}) as the intersection
	of the phase boundary between spin-ice and paramagnetic phase and the
	horizontal line with a specific ratio $T_\mathrm{peak}/D_\mathrm{nn}$.
An increase of $J_\mathrm{nn}$ by \SI{5}{\percent} from {\SI{-1.70+-.05}{\kelvin}}
	to {\SI{-1.79+-.05}{\kelvin}} can be found.
Therefore, the strength of the effective interaction
	$J_\mathrm{eff} = D_\mathrm{nn} + J_\mathrm{nn}$ is reduced from
	{\SI{0.69+-.03}{\kelvin}} to {\SI{0.63+-.03}{\kelvin}} by about
	{\SI{10}{\percent}}.
The reduction of the effective interaction is consistent with the reduced
	energy scale of the monopole excitation as seen in the ac
	susceptibility data
	(\autoref{fig:ac_susceptibility_phase_diagram}).
We find that the activation energy $\Delta \approx -9 J_\mathrm{eff}$ agrees
	well with previous measurements on {\dto{}} with
	$\Delta_{\dto{}} \approx -8.9 J_\mathrm{eff}$%
	{\citep{matsuhira_spin_2011, yaraskavitch_spin_2012}}.

The value of $J_\mathrm{nn}$ can also be obtained
	from comparison of the value of specific-heat peak $c_\mathrm{peak}$ at
	the peak temperature with theoretical
	calculations{\citep{den_hertog_dipolar_2000}}.
The experimental value of {$c_\mathrm{peak}$} is around {\SI{10}{\percent}}
	lower than would be
	expected from theory for an ideal spin-ice with
	{$J_\mathrm{nn}/D_\mathrm{nn} = -0.71$} or $-0.74$ obtained above,
	which might be due to crystal imperfections.
However, it is consistent with the value of previous measurements of the
	peak height of {\dgo{}}{\citep{zhou_chemical_2012}}.

After careful consideration of Ref.~{\citep{zhou_chemical_2012}},
	we come
	to the conclusion that the positioning of {\dgo{}} in the phase
	diagram ({\autoref{tab:phasediagram}}) of that work is at the lower
	end of the possible region.
While obtaining a similar value of $T_\mathrm{peak}$ they use a slightly
	higher value of $D_\mathrm{nn}$.
The fact that our {\dgsoiii{}} sample has a lower
	$J_\mathrm{nn}/D_\mathrm{nn}$
	than our {\dgo{}} as well as the sample investigated in
	Ref.~{\citep{zhou_chemical_2012}} proves
	that the reduction of the effective interaction in this compound is
	significant.

From these data, three conclusions are drawn.
\paragraph{Spin-ice characteristics of \dgsoxx{}:}
The XRD data show the high quality of the materials and confirm the pyrochlore
	structure with the \ce{Dy^3+} ions at the $A$ site and randomly
	distributed \ce{Ge} and \ce{Si} at the $B$ site.
The magnetization data are close to the expected curve of the powder-averaged
	paramagnetic Ising spins providing evidence for the Ising nature of the
	moments of the \ce{Dy^3+} ions due to the strong crystal electric field
	also observed in other spin-ice pyrochlores\citep{bertin_crystal_2012}.
The shape of the temperature-dependent ac susceptibility of \ce{Si}-substituted
	\ce{Dy} germanate resembles the ac susceptibility of the base compound
	and the well-studied spin-ice materials \dto{}%
	\cite{matsuhira_novel_2001} and \dso{}%
	\cite{matsuhira_low-temperature_2002}.
However, a substantial difference in the frequency dependence could be
	identified, which originates in a reduction of $\tau_0$.
The specific heats of \dgo{} and \dgsoiii{} have a similar shape at low
	temperatures also in accordance with previous measurements of the
	specific heat of the classical spin ices \dto{}%
	\citep{higashinaka_low_2004} and \hto{}%
	\citep{bramwell_spin_correlation_2001}.
Another evidence for {\dgsoiii{}} having spin-ice character is the residual
	entropy we observed.
However, the residual entropy of {\dgsoiii{}} is reduced compared to
	the Pauling entropy of the ideal spin-ice, possibly a side effect of the
	random distribution of {\ce{Si}} and {\ce{Ge}} on the $B$ site of the
	pyrochlore lattice.
In conclusion, the {\dgo{}} and {\dgsoiii{}} samples, representing {\dgsoxx{}},
	share several characteristic properties that are common among spin-ice
	materials.

\paragraph{Reduction of the effective nearest-neighbor interaction by
	\SI{10}{\percent}:}
\citet{zhou_chemical_2012} have found that substituting ions with smaller ionic
	radius on the $B$ site of the \ce{Dy}-pyrochlores \dso{} and \dto{}
	reduces the peak temperature of the magnetic specific heat.
The authors linked this to a reduction of the effective interaction due to the
	reduction of the distance between neighboring \ce{Dy^3+} ions using the
	$T_\mathrm{peak}/D_\mathrm{nn}$-$J_\mathrm{nn}/D_\mathrm{nn}$ phase
	diagram of the dipolar-spin-ice model\cite{den_hertog_dipolar_2000}.
Our XRD measurements confirm the reduced lattice constants in \dgo{} compared
	to \dto{}\citep{zhou_chemical_2012}.
The \ce{Si}-substituted samples continue this trend towards a reduction of the
	lattice constants and the peak temperatures in the specific heat as
	well as in the ac susceptibility.
Therefore, we argue, that in \dgsoxx{} the competing dipolar and exchange
	interactions are even further increased compared to \dgo{}.
Since the increase of the exchange interaction is stronger than the increase of
	the dipolar interaction, the total effective interaction is reduced
	compared to \dgo{}.

\paragraph{Influence of disorder on the spin-ice character}

A thorough study of the effects of disorder on the $B$ site of the spin-ice
	pyrochlores is not possible with the
	samples in this paper.
But still, the currently highly discussed influence of a stoichiometric or
	crystallographic disorder can be considered.
It is known{\citep{savary_disorder-induced_2017,martin_disorder_2017,
	wen_disordered_2017}} that
	strong modification of the regular structure can lead to the elimination
	of frustration, the formation of magnetically ordered clusters or the
	loss of the Ising character in pyrochlore compounds.
Quantum fluctuations can then have an increased effect and weaken the spin-ice
	character.
Although the distribution in the {\ce{Ge/Si}} system is random (could be
	checked by synchrotrons), the low {\ce{Si}} content does not seem to be
	sufficient to eliminate the magnetic frustration in the {\ce{Dy}}
	sublattice.
Averaging over microscopically different areas mostly recovers the spin-ice
	properties of the system.

\section{Summary}

High-quality polycrystals of the pyrochlores \dgsoxx{} with lattice constants
	down to \SI{9,906}{\angstrom} using the multi-anvil technique with
	pressures up to \SI{16}{\giga\pascal} have been synthesized.
The Ising nature of the moments of the \ce{Dy} ions was confirmed by the field
	dependence of the static magnetization.
Specific heat and ac susceptibility show the typical behavior of classical
	spin-ice compounds, namely a Schottky-like anomaly and residual entropy,
	and a frequency-dependent maximum, respectively.
The reduction of the lattice constant increases both the dipolar as well as the
	exchange-interaction strength.
However, while the dipolar interaction is only increased by less then
	\SI{1}{\percent}, the exchange interaction increases by
	\SI{5}{\percent}, leading to a reduction of the effective interaction
	by {\SI{10}{\percent}}.
Consequently, the silicon-substituted sample is closer to the phase boundary
	between the short-range spin-ice arrangement and the long-range AFM
	order.
The most significant difference between the substituted and unsubstituted
	compounds was found in the frequency dependence of the
	ac-susceptibility signal, which mainly originates in a reduction of the
	pre-exponential factor $\tau_0$.
However, a reduction of the energy scale of monopole excitations of
	\SI{1.5}{\percent} was observed as well, which underlines the finding
	of a reduced effective interaction.
In conclusion, we showed, that silicon substitution is a possible way to
	change the ratio of dipolar and exchange interaction and, hence,
	synthesize spin-ice compounds with customized properties.
The further increase of the substitutions on the $B$ site of the pyrochlores
	with improved high-pressure synthesis technology could open the
	possibility to study how disorder on the $B$ site of the pyrochlore
	influences the (spin-ice) characteristics%
	{\citep{savary_disorder-induced_2017}}, \eg{} through the formation of
	pinning centers, in view of work on other disordered materials%
	{\citep{li_crystalline_2017,zhu_disorder-induced_2017,
	sala_vacancy_2014}}.
Furthermore, the growth and investigation of single crystals of \dgo{} and
	its silicon-substituted compounds can be a challenging future
	task and offer the possibility to characterize the samples in terms of
	more recent spin-ice models including exchange interactions beyond
	nearest neighbors{\citep{mcclarty_chain-based_2015,
	henelius_refrustration_2016,rau_spin_2016}}.

\begin{acknowledgments}
We acknowledge support for using the ALBA synchrotron light source in Barcelona,
	Spain, and from HLD at HZDR, member of the European Magnetic Field
	Laboratory (EMFL).
This research has been supported by the DFG through SFB 1143
	(project-id 247310070).
\end{acknowledgments}

\bibliography{dissertation}

\end{document}